# Measuring Intangible Assets Using Parametric and Machine Learning Approaches[1]

Atika Nashirah Hasyyati and Adhi Kurniawan

BPS – Statistics Indonesia

December 6, 2022

**Abstract**

Intangible capital as the result of digitalization and globalization has not been fully measured yet in the economy because of several challenges. The limitation of data sources and the methodological issue related to how to measure and capitalize intangible assets are some fundamental issues. This paper aims at studying the contribution of intangible capital to business performance. The specific intangible capital, such as innovation, intellectual property, and branding are explored using parametric and machine learning methods. There are two data sources utilized in this study: survey data and Google Reviews data. Some variables are utilized as predictors based on the data sources. The variable selection techniques are implemented, followed by applying parametric regression and machine learning methods to predict business performance based on intangible capital variables. The results show that the proxy of intangible capital used in this paper has a significant contribution to business performance. In addition, variables that are obtained from google reviews can be used to predict the use of branding with high accuracy.

---

[1] Presented in the 10th IMF Statistical Forum: Measuring the Tangible Benefits of Intangible Capital, 17 November 2022

## I. Introduction

Intangible capital as the result of digitalization and globalization has not been fully measured yet in the economy because of several challenges. The limitation of data sources and the methodological issue related to how to measure and capitalize intangible assets are some fundamental issues. However, most countries need to calculate the contribution of intangible capital in the digital economy era.

"The digital economy refers to a broad range of economic activities that include using digitized information and knowledge as the key factor of production, modern information networks as an important activity space, and the effective use of information and communication technology (ICT) as an important driver of productivity growth and economic structural optimization" (G20, 2016). Meanwhile, OECD (2020) defines the digital economy as all economic activity reliant on, or significantly enhanced by the use of digital inputs, including digital technologies, digital infrastructure, digital services, and data. It refers to all producers and consumers, including the government, that are utilizing these digital inputs in their economic activities. However, Saunders & Brynjolfsson (2016) suggest that the contributions of IT to value depend heavily on other factors, and are not a rising tide that lifts all boats.

Intangible capital is defined as assets that have no physical or financial embodiment (Alsamawi et al., 2020). Corrado et al. (2005) classified intangible assets into three categories: (i) computerized information; (ii) scientific and creative property; and (iii) economic competencies. Vosselman (1998) classified intangible investment into two components: core components and supplementary components. The core components consist of research and experimental development (R&D); education and training; software; marketing; mineral exploration; licenses, brand, and copyright; and patents. The supplementary categories of intangible investment are the development of an organization, engineering and design, construction and the use of databases, remuneration and innovative ideas, and other human resource development (training excluded).

In Oslo Manual (2018), it is stated that an innovation is a new or improved product or process (or a combination thereof) that differs significantly from the unit's previous products or processes and that has been made available to potential users (product) or brought into use by the unit (process). Furthermore, market research and testing, pricing strategies, product placement, and product promotion are all aspects of marketing and brand equity activities. Other marketing and brand equity activities include product advertising, promoting products at trade shows or exhibitions, and developing marketing strategies. The protection or utilization of knowledge, frequently developed through R&D, software development, engineering, design, and other creative efforts, is included in intellectual property-related activities. Meanwhile, the term "intellectual property activities" refers to all administrative and legal tasks involved in obtaining, registering, documenting, managing, trading, licensing-out, marketing, and enforcing a company's own intellectual property rights (IPRs), as well as all actions taken to obtain IPRs from other businesses, such as through licensing-in or outright IP acquisition, and actions taken to sell IP to third parties.

Regarding IT capital, Byrne (2022) states that there are several other noteworthy changes in the nature of digitalization have accompanied the shift in industrial composition of IT capital investment: (1) increasing reliance on purchased IT services; (2) radical increase in mobility; (3) shift toward intangible investment; (4) ongoing explosion of data.

This paper aims at studying the association between intangible capital and business performance using survey data. We also examine the accuracy of intangible capital variables to predict business performance using some statistical methods. In addition, utilizing google review data, we investigate the potential effect of information in the google review for branding.

## II. Measuring Intangible Capital based on Survey Data

**Methodology**

The first data source is the 2021 Business Characteristics Survey (BCS). The BCS is a probability survey conducted annually in all provinces in Indonesia to provide the estimation at a national level. The target sample size is 8.300 medium and large enterprises. The data is collected through a face-to-face interview between the enumerator and the respondent. The BCS covers intellectual property rights ownership; benefits from owning intellectual property rights; business use of ICT indicators; business activities conducted using the internet such as video conferencing, e-commerce, the use of social media and instant messaging, e-government, internet banking, accessing other financial facilities, digital products deliveries, employment recruitments, training; website ownerships; information on innovation conducted by businesses, including the percentage of expenditure for innovation compared to total expenditure.

Income is used as a response variable. The total number of predictors is six variables. The predictors consist of four variables representing intangible capital (intellectual property, IT infrastructure, the existence of a website for promotion/branding, and product innovation). Moreover, two supplementary independent variables are not directly related to intangible capital including in the model (the number of staff/employees and foreign investment).

**Descriptive Statistics**

Based on the 2021 Business Characteristics Survey data, we examine the relationship between the utilization of intangible capital and the income obtained by enterprises. In general, medium and large enterprises using intangible capital tended to have higher incomes compared to enterprises that did not utilize intangible assets. For enterprises using intangible capital, the number of enterprises having a high income (more than the logarithm of income) was larger than the number of enterprises having a low income. On the contrary, in the case of enterprises that did not use intangible capital, the number of enterprises having a low income (less than the logarithm of income) was larger than the number of enterprises having a high income.

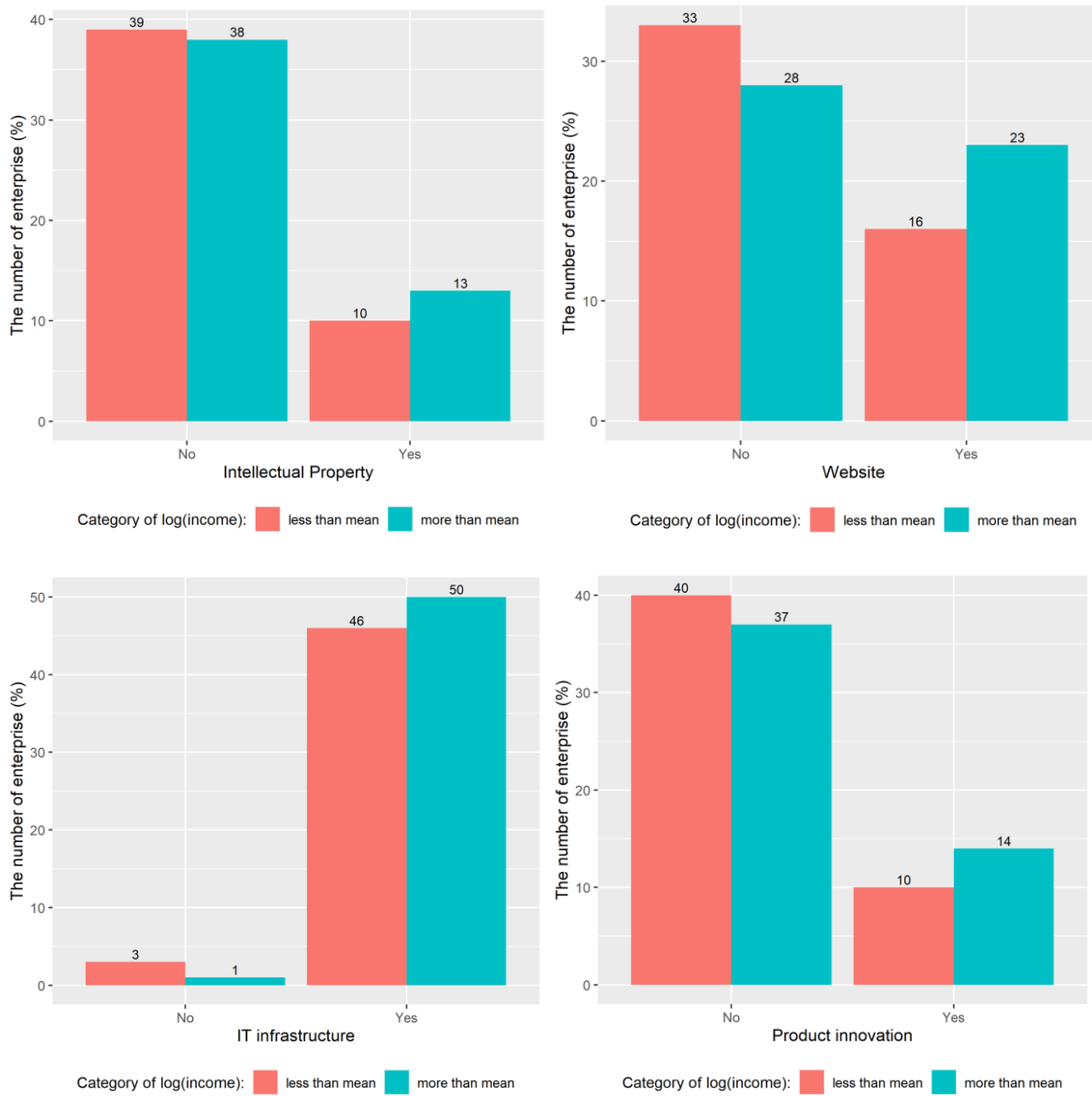

Figure 1. The number of enterprises (%), split by the use of intangible capital and log(income)

The utilization of IT infrastructure was relatively high. It was estimated that approximately 96% of medium and large enterprises use IT infrastructure in their business activities. However, only 39% of enterprises had a website for branding strategy. In terms of intellectual property, the number of enterprises having intellectual property was relatively low, around 23% of the total enterprise. Moreover, the percentage of enterprises that had product innovation was also low in percentage (about 24%).

**Parametric Approach**

Initially, we run a parametric regression model using survey weight to examine the association between some related to intangible capital and the income of the company. Survey weight is taken into account for building the model since the data is collected by using *unequal probability sampling*. Moreover, a log transformation of the dependent variable is chosen since the data of income exhibit right skewness (positively skewed).

The model is defined as:

$$\log(y) = \beta_0 + \beta_1 x_1 + \beta_2 x_2 + \beta_3 x_3 + \beta_4 x_4 + \beta_5 x_5 + \beta_6 x_6 + \varepsilon$$

$y$ : income

$x_1$ : the number of employees

$x_2$ : foreign investment (0=No, 1=Yes)

$x_3$ : intellectual property (0=No, 1=Yes)

$x_4$ : IT infrastructure (0=No, 1=Yes)

$x_5$ : the use of the website for branding (0=No, 1=Yes)

$x_6$ : product innovation (0=No, 1=Yes)

Table 1. Estimates of the regression coefficient

| Variable | Estimate | Standard Error | t.value | p-value |
|---|---|---|---|---|
| Intercept | 20.201 | 0.139 | 145.63 | 0.00 |
| The number of employees | 0.001 | 0.000 | 1.59 | 0.11 |
| Foreign investment | 1.878 | 0.198 | 9.48 | 0.00 |
| Intellectual property | 0.055 | 0.075 | 0.73 | 0.47 |
| IT infrastructure | 1.197 | 0.144 | 8.32 | 0.00 |
| Website for branding | 0.597 | 0.066 | 9.02 | 0.00 |
| Product innovation | 0.360 | 0.077 | 4.69 | 0.00 |

Based on Table 1, it is obvious that most intangible capital variables are statistically significant in the model at a 5% of the significance level. IT infrastructure has the highest regression coefficient among the intangible variables, indicating that it has the largest significant association with the income gained by the enterprise. The use of the website for branding and product innovation is also significant in the model, while intellectual property is not significant and has the lowest regression coefficient estimate among the variables related to intangible capital.

Based on the result of regression coefficient estimates, the average difference in income between businesses that used intangible capital and businesses that did not use intangible capital can be obtained. It is estimated that the income of businesses having IT infrastructure is 3.32 times higher than the income of businesses that did not use IT infrastructure (among observations with the same values of all the other predictors). In addition, the income of businesses that used the website for branding/promotion is 82% larger than the income of businesses that did not use the website (among observations with the same values of all the other predictors). Furthermore, businesses with product innovation tend to have income 1.43 times higher than those who did not have innovation (among observations with the same values of all the other predictors).

**Prediction**

Instead of inferential purpose (estimating regression coefficients), another aim of building a parametric regression model is to predict the response variable. In this section, the comparison of accuracy between four methods to predict the log(income) is based on six predictors. The methods that are used for prediction are parametric regression with logarithm transformation, KNN regression, Bootstrap Aggregation (Bagging), and Random Forest.

K-Nearest Neighbours (KNN Regression) is a nonparametric method that can be used to predict a response variable by averaging the observations of k-nearest neighbors. In general, prediction using Bagging involves three main steps: selecting bootstrap samples, fitting a model to each sample, and averaging the prediction from each sample. Random forest procedure is similar to the Bagging procedure, but there is some modification. The random forest procedure is commenced by drawing the bootstrap sample. Then, for each sample, we fit a tree from some specified depth. At each split, we select the splits using only a randomly selected subset of some variables, rather than choosing from all variables. Ultimately, we average the prediction using only the "out of bag" (OOB) samples. Meanwhile, XGBoost is a scalable machine-learning system for tree boosting. A cross-validation procedure is conducted to evaluate the performance of these methods for prediction by estimating the test error.

Table 2. The comparison of prediction accuracy

| No | Method | RMSE | R-Square | MAE |
|---|---|---|---|---|
| 1 | Parametric regression | 2.3508 | 0.0854 | 1.7528 |
| 2 | KNN-regression | 1.9640 | 0.3224 | 1.4646 |
| 3 | Bagging | 1.9502 | 0.3293 | 1.4461 |
| 4 | Random Forest | 1.9500 | 0.3418 | 1.4519 |
| 5 | XGBoost | 1.9331 | 0.3471 | 1.4347 |

Looking at Table 2, predictions based on the machine learning approach tend to have higher accuracy than the parametric approach. Predicting log(income) using XGBoost results in the lowest Root Mean Square Error (RMSE), the highest R-square, and the lowest Mean Absolute Error (MAE). On the other hand, there is just a slight difference between the accuracy of KNN-regression, Bagging, and Random Forest.

### III. Measuring Intangible Capital based on Google Reviews Data

**Google Reviews Data Collection and Processing**

In collecting Google Reviews data set through Google Places API ($200 free for the first month of using Google Cloud Platform), there were several steps conducted in this study.

Firstly, using hotel-related keywords in Indonesia as a study case of the use of branding in businesses. There is a limitation in gathering the data by using the free $200 from Google Places API to only 20 hotels per keyword. Therefore, in this study, we conducted research by using search terms, such as "star hotels in (name of each province or city in Indonesia)" and the Indonesian language "hotel berbintang di (name of each province or city in Indonesia)". Based on 34 provinces in Indonesia, we collected 1622 hotels from Google Places API.

The second step was data wrangling. Data wrangling took approximately 98% of the time needed in the overall data processing. The data wrangling process also comprises recoding some variables, such as hotels with photos, types (whether there is information related to facilities provided by hotels), and opening hour information on Google Reviews. In addition, we have to do deterministic matching of data frames in each province for 34 provinces since the limitation from the free Google Reviews data is 20 observations that we collected into three data frames on average in each province. This process aims at removing duplicate observations from the database. After that, data frames were combined with the prerequisite the data frames need to be in the same structure.

| | business_status | formatted_address | geometry | icon | icon_background_color | icon_mask_base_uri | name | photos | place_id | plus_code |
|---|---|---|---|---|---|---|---|---|---|---|
| 116 | OPERATIONAL | Jl. Kuala Simeme, pa... | NA | https://maps.gsta... | #909CE1 | https://maps.gstatic.co... | Pancur Gading H... | 0 | ChIJiwf78GYwMT... | NA |
| 117 | OPERATIONAL | Jl. Sisingamangaraja ... | NA | https://maps.gsta... | #909CE1 | https://maps.gstatic.co... | Antares Hotel | 0 | ChIJDdMcTF0wM... | NA |
| 118 | OPERATIONAL | Commercial & BizPar... | NA | https://maps.gsta... | #909CE1 | https://maps.gstatic.co... | The Lively Hotel ... | 0 | ChIJT2uD7CE2M... | NA |
| 119 | OPERATIONAL | 2, Jl. Komp. Pelindo l... | NA | https://maps.gsta... | #909CE1 | https://maps.gstatic.co... | hotel | 0 | ChIJMXnZvWshai... | NA |
| 120 | OPERATIONAL | Jl. Gereja No.5, RW.5,... | NA | https://maps.gsta... | #909CE1 | https://maps.gstatic.co... | Hotel | 0 | ChIJbWajrP_1aS4... | NA |
| 121 | OPERATIONAL | VQ8V+465, Jakarta ... | NA | https://maps.gsta... | #909CE1 | https://maps.gstatic.co... | Hotel | 0 | ChIJ8VrHAZwdai... | NA |
| 122 | OPERATIONAL | Kompleks, Jl. Komp. ... | NA | https://maps.gsta... | #909CE1 | https://maps.gstatic.co... | Star Inn | 0 | ChIJp_hc66kxMT... | NA |
| 123 | OPERATIONAL | Jl. Prof. H. M. Yamin ... | NA | https://maps.gsta... | #909CE1 | https://maps.gstatic.co... | Cordela Hotel M... | 1 | ChIJmcErvb4xMT... | NA |
| 124 | OPERATIONAL | Jl. Sultan Serdang No... | NA | https://maps.gsta... | #909CE1 | https://maps.gstatic.co... | Prime Plaza Hot... | 1 | ChIJ69MoeR82M... | NA |
| 125 | OPERATIONAL | Jl. Gagak Hitam No.1... | NA | https://maps.gsta... | #909CE1 | https://maps.gstatic.co... | Saka Hotel Medan | 1 | ChIJu9XRPzr0aS4... | NA |
| 126 | OPERATIONAL | Jl. Sisingamangaraja ... | NA | https://maps.gsta... | #909CE1 | https://maps.gstatic.co... | Hotel Menara Le... | 1 | ChIJOc6bSF0wM... | NA |
| 127 | OPERATIONAL | Jl. S. Parman No.217,... | NA | https://maps.gsta... | #909CE1 | https://maps.gstatic.co... | Cambridge Hote... | 1 | ChIJx46Jw9MxM... | NA |
| 128 | OPERATIONAL | Jl. Jend. Ahmad Yani ... | NA | https://maps.gsta... | #909CE1 | https://maps.gstatic.co... | Kama Hotel | 1 | ChIJuRIrQbUxMT... | NA |
| 129 | OPERATIONAL | HRXJ+452, Jl. Bandar... | NA | https://maps.gsta... | #909CE1 | https://maps.gstatic.co... | The Crew Hotel | 1 | ChIJzWdwISA2M... | NA |

| rating | reference | types | user_ratings_total | opening_hours | prov |
|---|---|---|---|---|---|
| 4.4 | ChIJiwf78GYwMTARZGbjbCiQB-w | 0 | 701 | 0 | North Sumatra |
| 3.9 | ChIJDdMcTF0wMTARqIxz7joYopU | 0 | 812 | 0 | North Sumatra |
| 4.0 | ChIJT2uD7CE2MTAR-s8UrXQWaAs | 0 | 469 | 0 | North Sumatra |
| 0.0 | ChIJMXnZvWshai4R_QuKk0xp7F4 | 0 | 0 | 0 | Jakarta |
| 4.0 | ChIJbWajrP_1aS4RJnRKMHw_o9Y | 0 | 1 | 0 | Jakarta |
| 2.0 | ChIJ8VrHAZwdai4RTcxtO9dE110 | 0 | 1 | 0 | Jakarta |
| 0.0 | ChIJp_hc66kxMTARt62fIbJuGTU | 0 | 0 | 0 | North Sumatra |
| 4.2 | ChIJmcErvb4xMTARVrDutkAz3H0 | 1 | 1202 | 1 | North Sumatra |
| 4.4 | ChIJ69MoeR82MTAR570-5FjDTzU | 1 | 1722 | 1 | North Sumatra |
| 4.1 | ChIJu9XRPzr0aS4RY4LyoCTGOTQ | 1 | 2188 | 1 | North Sumatra |
| 3.8 | ChIJOc6bSF0wMTARQ2hpIqCItns | 1 | 645 | 0 | North Sumatra |
| 4.6 | ChIJx46Jw9MxMTARU7Qg4PCydrQ | 1 | 4375 | 1 | North Sumatra |
| 4.1 | ChIJuRIrQbUxMTAR4Oyzk-kpLcc | 1 | 1437 | 1 | North Sumatra |
| 4.3 | ChIJzWdwISA2MTARPvSbmA-k4QQ | 1 | 334 | 0 | North Sumatra |

Figure 2. Example of Google Reviews Data Set

**Machine Learning Approach**

Firstly, an unsupervised machine learning method was used to get categories of branding based on the Google Reviews data. We use six variables obtained from google review data: (1) the availability of photos, (2) rating, (3) the availability of information about facilities, (4) the number of reviewers, (5) opening hour information, (6) province. Since there are variables with mixed data types, the concept of Gower distance with Partitioning Around Medoids (PAM) is implemented. In selecting the number of clusters, we applied Silhoutte width which is an internal validation metric as an aggregated measure of how similar an observation is to its own cluster compared to its closest neighboring cluster. In this case, the number of clusters is two clusters according to the highest Silhouette width.

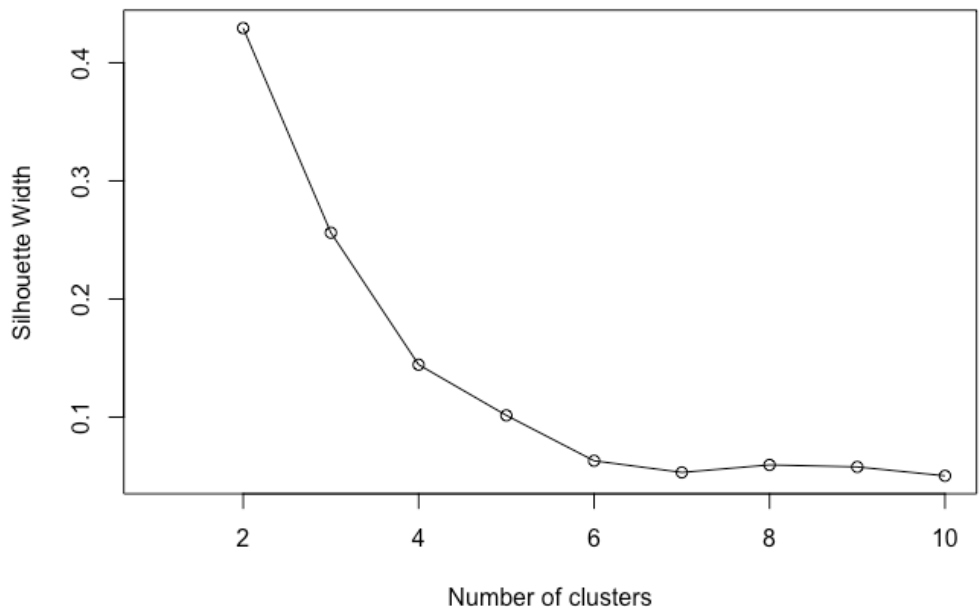

Figure 3. The Number of Clusters

Using t-distributed stochastic neighborhood embedding (t-SNE), the following plot preserves the local structure of the variables in a lower dimensional space to make the clusters visible. The plot shows that there are two well-separated clusters according to the PAM method. However, there are thirty observations with distinct characteristics with the two clusters (not enough observations to make another cluster).

There are 897 observations in cluster 1. Most of the observations that are in the first cluster provide information about facilities, photos, and opening hours. On average, those observations have 2539 total user ratings and the average rating is 4.3. Meanwhile, there are 725 observations in cluster 2. The characteristics of observations in cluster 2 are similar to observations in cluster 1 in the case that most of them provide information about facilities and photos. However, only 7 of them provide information about opening hours and the average user rating total is 1581 with an average rating is 4.0.

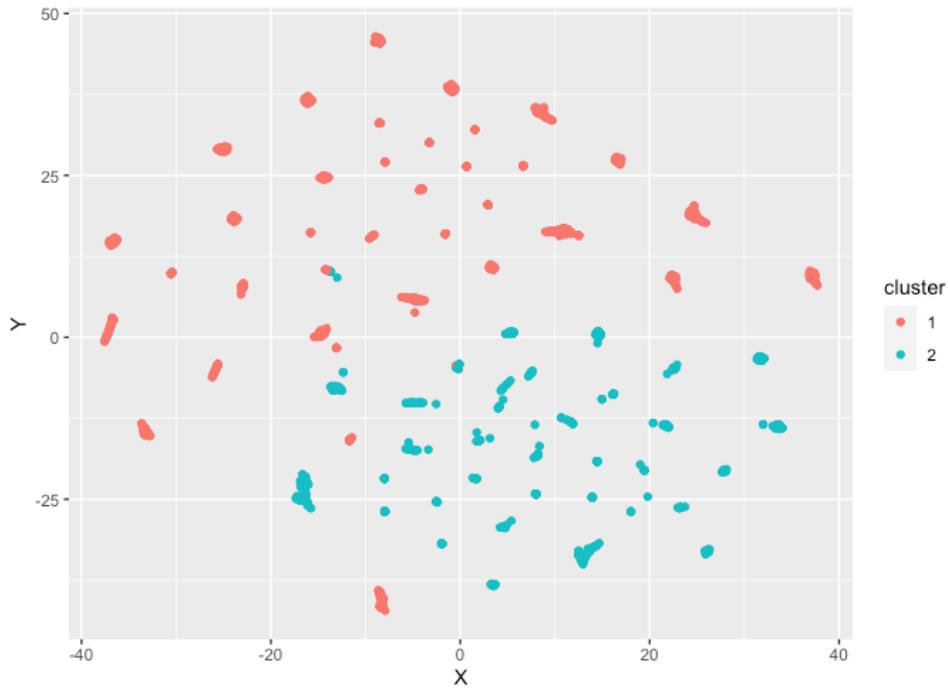

Figure 4. Visualization using t-distributed stochastic neighborhood embedding (t-SNE)

In order to examine the reassure that those variables are good predictors for branding category, some supervised learning methods are performed, such as regression trees, random forest, neural network, XGBoost, and penalized multinomial regression. All methods result in the same conclusion that the accuracy is very high (around 99%).

Table 3. The comparison of prediction accuracy

| No | Method | Accuracy | Kappa |
| --- | --- | --- | --- |
| 1 | Regression tree | 0.9938 | 0.9875 |
| 2 | Random forest | 0.9977 | 0.9953 |
| 3 | Neural network | 0.9969 | 0.9937 |
| 4 | XGBoost | 0.9992 | 0.9984 |
| 5 | Penalized multinomial regression | 0.9984 | 0.9968 |

**Association between rating and the number of reviewers**

Based on the google reviews data and official survey data, we can estimate the average rating at the province level and the ratio of reviewers to visitors using these formulas:

$$I_i = \frac{P_{ij}I_{ij}}{\sum_j P_{ij}I_{ij}}$$

where:

$I_i$ : the rating average at *i*-th province

$P_{ij}$ : the number of reviewers at *i*-th province *j*-th hotel

The ratio of reviewers to visitors is estimated by:

$$r_i = \frac{\frac{N_i}{n_i}\sum_j P_{ij}}{O_i T_i}$$

$r_i$ : ratio of the number of reviewers to the number of visitors in *i*-th province

$O_i$ : occupancy rate in *i*-th province (based on official survey data)

$T_i$ : the total number of hotel rooms in *i*-th province (based on official survey data)

$N_i$ : the total population of hotels in *i*-th province (based on official survey data)

$n_i$ : the number of hotels in the dataset (based on google review data) in *i*-th province

We estimate $I_i$ and $r_i$ for each province, and examine the association between them.

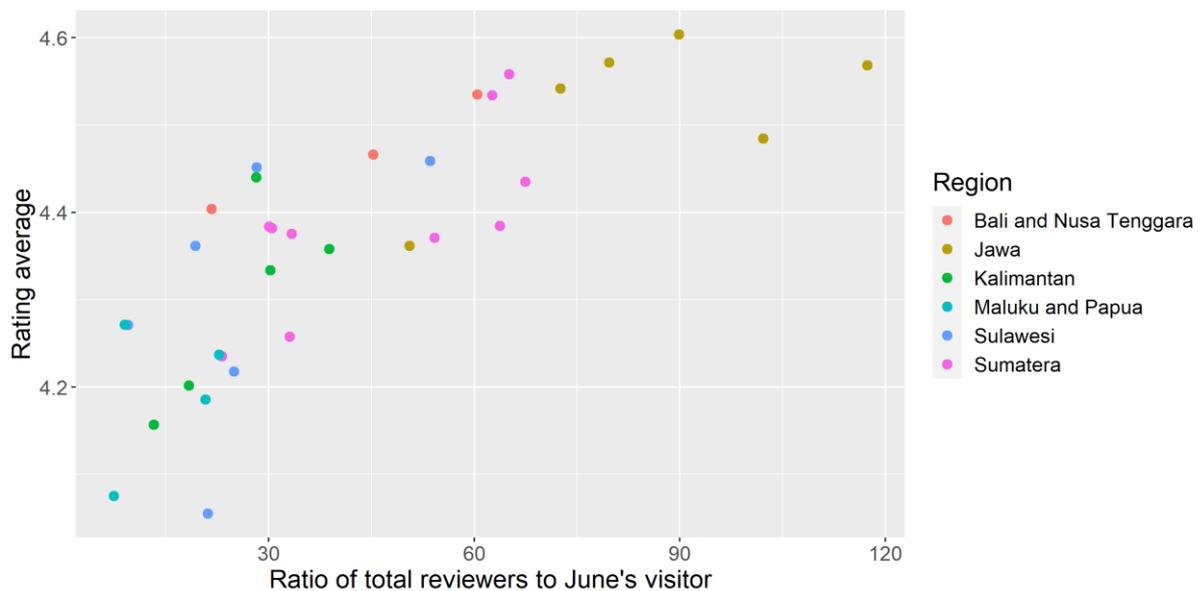

Figure 5. Scatter plot between rating average and ratio of total reviewers to June's visitor

Figure 5 describes the association between rating and the number of reviewers. There is a positive linear correlation between the rating and the number of reviewers with a Pearson correlation coefficient of approximately 0.78. The higher the number of reviewers, the higher the rating average of the province. It also indicates that visitors who wrote the review tend to be the visitors who had a good experience when staying in the hotel. Hence, Google Reviews will be a potential asset for the branding strategy of the accommodation. The tendency of hotel visitors to give reviews in Jawa (Java) is higher than those on other islands.

## IV. Conclusion

The results of parametric regression show that the proxy of intangible capital used in this paper, except intellectual property, has a significant contribution to business performance. In addition, based on the machine learning approach, variables that are obtained from Google Reviews can be used to predict the use of branding with high accuracy. We recommend that the combination of using official survey data and other sources of data (e.g. Big data) in statistical analysis through statistical data integration should be continuously developed to measure intangible capital.